\documentstyle[referee]{l-aa}
\begin{document}
\thesaurus{02(02.13.2; 02.18.8; 12.12.1)}
\title{Magnetic fields and large-scale structure in a hot
universe.}
\subtitle{II. Magnetic flux tubes and filamentary structure}
\author{E. Florido \and E. Battaner}
\institute{Dpto. Fisica Teorica y del Cosmos. Universidad de Granada.
Spain.}
\maketitle

\begin{abstract}
In paper I, we obtained an equation for the evolution of density
inhomogeneities in a radiation dominated universe when they are affected
by magnetic fields. In this second paper we apply this equation to the case
in which the subjacent magnetic configuration is a flux tube. For scales
of the order of 1 Mpc or less the differential equation is elliptical. 
To solve it, we have used the numerical method based on  "Simultaneous Over
Relaxation", SOR, with Chebyshev acceleration and we have treated
the problem as a boundary value problem, which restricts the prediction
ability of the integration. For large-scale flux tubes, much larger than
1 Mpc, the equation can be analytically integrated and no assumption
about the final shape or magnitude of the inhomogeneity is required. 
In both cases we obtain an evolution which does not differ very much from
linear in time. The inhomogeneity in the density becomes filamentary. Large
scale structures ($\ge$ 10 Mpc) are probably unaffected by damping,
non-linear and amplification mechanisms after Equality, so that this model
provides a tool to interpret the present observed large scale structure.
 Filaments are very frequently found in the large-scale structure in the Universe. It is suggested
here that they could arise from primordial magnetic flux tubes, thus providing
an alternative hypothesis for its interpretation; in particular we
consider the case of the coma-A1367 supercluster, where the magnetic field
is known to be high.


\end{abstract} 

\section{General equations}
In paper I, we presented the basic equations governing the evolution
of the large scale energy density structure in the prerecombination
universe, when the effect of magnetic fields  cannot be ignored. In the
linear regime, the magnetic field configuration remains time-independent
throughout this era, only growing with the expansion. The degree and the way
this field structure affects the evolution of density inhomogeneities
would depend on the particular pattern of the magnetic field lines. We
will consider here a small cell in which the magnetic field has the
simplest structure: a magnetic flux tube. Different sizes of the flux tube
will be considered, for small, intermediate and large scales.

We will deal with the integration of equation (86) of paper I

\begin{center}
$\ddot{\delta} -\delta -X +{1 \over 3} e^{-\tau} \nabla'^{2} \delta
+2 e^{-\tau} m =0$\\
\end{center}
where:\\
$\delta$ is the relative energy density contrast, defined as $\delta
\epsilon/\epsilon$, where $\epsilon$ is the energy density and
$\delta \epsilon$ the difference between its value within the inhomogeneity
and its mean value in the Universe. As we are considering relativistic
particles, we also have $\delta = \delta p/p$ where $p$ is the hydrostatic
pressure of the relativistic particles (either photons or any hot dark
matter particles).

$\ddot{\delta}$ is the second derivative with respect to the time-like
variable $\tau$.

$\tau$ is defined as $\tau = \ln{(t_e/t)}$ where $t$ is the time and
$t_e$ is the last time considered in this paper, close to equality.
More specifically we have considered $R_e = 10^{-5}$, or $t_e =3.66
\times 10^{9} s$, before the acoustic epoch.

$\nabla'$ is the  Laplacian operator, when the spatial coordinates
are $x'_i$ instead of $x_i$, the usual comoving coordinates.

$x'_i$ is defined as $x'_i=(K/R_e)x_i$, where $K$ is the constant in the expansion
law $R^2=Kt$. The value of $K$ is $2.73 \times 10^{-20}s^{-1}$ and therefore
$x'_i = 2.73 \times 10^{-15}x_i$, where $x_i$ are measured in seconds.
The relation between $x'_i$ and $d$, the length measured in present-day--Mpc is
$x'_i =0.28(d/Mpc)$.

$X$ characterizes the magnetic energy density and is defined as
\begin{center}
 $X = {{B_0^2} \over {24 \pi p_0}}$
\end{center}
\indent ${\vec{B}_0}$ would be the present magnetic field if no source and no
loss other than expansion had taken place after the time period
considered here. Because the post-recombination epoch is probably
very complicated, concerning the evolution of magnetic fields, $\vec{B}$
has in practice no relation with present magnetic fields. $\vec{B}_0$ is
exactly defined as $\vec{B}_0 = \vec{B} R^2$. We measure magnetic field
strengths in $s^{-1}$, with the equivalence being 1 Gauss =$8.61 \times
10^{-15} s^{-1}$.

$p_0$ is the relativistic particle hydrostatic pressure at present. It
is defined as $p_0 =pR^4$ and we have chosen the value $p_0 = 8.84
\times 10^{-42}s^{-2}$.

$m$ also characterizes the magnetic field configuration
\begin{center}
  $m=-{1 \over 2} \nabla' \cdot \left( {2 \over 3} \vec{n} - \nabla' X \right)$
\end{center}
where
\begin{center}
  $\vec{n} =-{{\vec{B}_0 \cdot \nabla \vec{B}_0} \over {4 \pi p_0}}
           +3 \nabla' X$
\end{center}
To integrate our basic equation it is then necessary to specify the field.
In this paper we analyze the influence of a magnetic tube flux, i.e.
$\vec{B}_0$ is given by
\begin{equation}
  \vec{B}_0 = (0,0,A) e^{-{{r^2} \over {2\sigma^2}}}
\end{equation}
where $r$ and $\sigma$ are measured with the same unit as $x'_i$. For
instance $\sigma =1$ is equivalent to a comoving-length of 0.28 Mpc.
For our purposes, the solution very much depends on the value of $\sigma$
and we call it large scale if $\sigma \gg 1$, intermediate scale, if
$\sigma \approx 1$, and small scale, if $\sigma \ll 1$.

In the particular case of a flux tube:
\begin{center}
  $X = a e^{-{{r^2} \over {\sigma^2}}}$
\end{center}
where
\begin{center}
  $a= {{A^2} \over {24 \pi p_0}}$
\end{center}
and
\begin{center}
  $m={{2a} \over {\sigma^2}} e^{-{r^2 \over \sigma^2}} \left(
     2-{r^2 \over \sigma^2} \right)$
\end{center}
For integrating, it is better to define another time-like variable as
\begin{center}
  $t' = e^{-\tau} = {t \over t_e}$
\end{center}

Using this dimensionless time the basic equation becomes:
\begin{equation}
  \delta'' t'^2 + \delta' t' -\delta -X+{1 \over 3}\hskip 0.5mm t'
  \hskip 0.5mm \nabla'^2 \delta
  +2\hskip 0.5mm t'm=0
\end{equation}
where now $\delta'$ and $\delta''$ are first and second derivatives
with respect to $t'$.

From now on we will write simply $t$ instead of $t'$ and $x_i$
instead of $x'_i$ without any risk of confusion.

\section{The integration and boundary conditions}

Equation (2) is an elliptic linear second order differential equation with
variable coefficients. From the physical point of view, it would be
preferable to carry out the integration as an initial value problem,
to start with an initial configuration and calculate the shape and
density of the cloud at the final step, at $t_e$. However, elliptic
differential equations cannot usually be integrated this way, and
a unique solution does not exist. Our first attempts to treat the problem
as an initial value problem were indeed very unstable, confirming this fact.
Probably, this intrinsic instability of the equation is somewhat
associated with a physical complexity. It was therefore necessary to
look for a boundary value solution. In this case, a unique solution does exist,
but as we must assume the final geometry of the cloud, the predictive
possibilities are completely lost. Nevertheless, the evolution of the
cloud can be followed by means of very stable methods and some combinations
of free parameters and boundary conditions can be rejected if they
provide physically implausible solutions.

We chose  "Simultaneous Over-Relaxation" method (SOR) with 
Chebyshev acceleration (Press et al. 1989; Holt, 1984; Smith, 1985
and others).

The equation is written as

 $\begin{array}{l} 
  \delta_{l+1,j}\left( l^2+{l \over 2} \right) +
  \delta_{l-1,j} \left(l^2 -{l \over 2} \right) +
  \delta_{l,j+1} \left({{l \Delta t} \over {3(\Delta x)^2}} +
    {{l\Delta t}\over {6j(\Delta x)^2}} \right) +\\
~~~~~~~~~~~\\
  \delta_{l,j-1} \left({{l \Delta t} \over {3(\Delta x)^2}} -
    {{l\Delta t}\over {6j(\Delta x)^2}} \right) +
  \delta_{l,j} \left(-2l^2 -1 -{{2l\Delta t} \over {3(\Delta x)^2}} \right)-\\
~~~~~~~~~~~~~\\
  a e^{-j^2(\Delta x)^2/\sigma^2}-
  {{4al\Delta t} \over \sigma^2} e^{-j^2(\Delta x)^2/\sigma^2}
    \left({{j^2(\Delta x)^2} \over \sigma^2} -1 \right) =0
 \end{array}$

or equivalently
\begin{center}
  $a_{l,j} \delta_{l,j+1} +b_{l,j} \delta_{l,j-1}+ c_{l,j}\delta_{l+1,j} 
  +d_{l,j} \delta_{l-1,j} +e_{l,j} \delta_{l,j} =f_{l,j}$ 
\end{center}
with obvious definitions of the coefficients $a_{l,j}$, $b_{l,j}$, 
$c_{l,j}$, $d_{l,j}$, $e_{l,j}$ and $f_{l,j}$. Subindex $l$ denotes
time and subindex $j$ the spatial variable $r$. The iterative method
calculates a new $\delta$ map from a previous step $\delta$ map by
\begin{center}
  $\delta_{l,j}^{new} = \delta_{l,j}^{old} -\omega {\xi_{l,j} \over e_{l,j}}$
\end{center}
where $\xi$ is the residual calculated by
\begin{center}
  $\xi_{l,j} = a_{l,j} \delta_{l,j+1} +b_{l,j} \delta_{l,j-1}+ c_{l,j}\delta_{l+1,j}
  +d_{l,j} \delta_{l-1,j} +e_{l,j} \delta_{l,j} -f_{l,j}$
\end{center}
and $\omega$ is the relaxation parameter. When using Chebyshev acceleration,
$\omega$ is estimated at every iterative step. The network is divided
into white and black points as in a chess-board. The value of $\delta$
in white points is calculated from the previous step values of $\delta$ in
black points, and at the next step $\delta$ in black points are 
calculated from $\delta$ in white points. The relaxation parameter is
calculated with the series\\
\begin{center}
$ \omega^{(0)} =1$\\
$\omega^{(1/2)} = 1/\left(1-\rho_{Jacobi}^2 /2 \right)$ \\
$....$\\
$\omega^{(n+1/2)}=1/\left( 1-\rho_{Jacobi}^2 \omega^{(n)}/4 \right)$\\
$....$\\
$\omega^{(\infty)}=\omega_{optimus}$
\end{center}
where\\
\begin{center}
$\rho_{Jacobi} = {{\cos{\pi \over J} +\left({{\Delta x} \over {\Delta t}}\right)^2
  \cos{\pi \over L}} \over {1+ \left({{\Delta x} \over {\Delta t}}\right)^2}}$
\end{center}
if the size of the net is $L \times J$.

Convergence was usually obtained in less than 800 steps, and the solution is
very stable.

We need to take boundary conditions in space and in time (see Figure 1). Far from the
flux tube (at about $\pm 3\sigma$) we would have $\delta =0$, for instance
for $r=3\sigma$. The other space-boundary could be either a van 
Neuman condition, $\dot{\delta} (r=0)=0$, or again $\delta (r=3\sigma)=0$
in the opposite direction. Because of the symmetry of the flux tube
they must be equivalent, with the latter condition being more time and memory demanding.
We have tried both and obtained the same result. This was one way to test
the stability of the SOR.

\begin{figure}
\caption[]{Diagram of the boundary conditions}
\end{figure}

With respect to the time-boundaries, we have at $t=0$ two possibilities.
Either $\delta(t=0)=0$, which we call "homogeneity" or $\delta (t=0)=-X$, which
we call "isocurvature". In the first case, it is implicitly assumed
that no inhomogeneities are initially present: these are subsequently produced
by magnetic field structures. As the presence of magnetic fields introduces
a metric perturbation, the isocurvature condition assumes that this energy
density excess is initially compensated by an under-concentration of the 
dominant particles, so that the curvature is initially constant. We have
numerically found that the two conditions produce different behaviours only
in the very first time steps and that the evolution coincides through most
of the period considered. This is discussed below. We have not begun
at $t=0$ exactly but at $t=0.01$ (remembering that $t$ varies between 0
and 1). On the one hand $t=0$ may introduce some instability, as discussed below,
as foreseen in the theory. On the other hand $t=0$ is Big
Bang time, which is beyond our scope.

At $t=1$, we have adopted $\delta (t=1)=\epsilon X$, i.e. with $\delta(r)$
being a gaussian. The parameter $\epsilon$ was adopted such that $\delta 
(t=1,r=0)$ was $10^{-4}$, because this would be a typical value of $\delta$
at $t_e$, in order to reproduce the present inhomogeneity field.

Results for low and intermediate scales were numerically found by the
above described procedure. For the larger scales, it is shown later
that the solution can be theoretically found.

\section{Small and intermediate flux tube thickness}

We have two basic parameters: $\sigma$, which determines the length-scale
of the inhomogeneity, and $a$, which determines the magnetic field strength.
After some initial trial calculations we decided to adopt $a$ in the range
$10^{-5}$ to $10^{-4}$. A value much lower than this makes the problem
a classical one in the absence of magnetic fields. A higher value produced
unrealistic profiles with a large maximum at intermediate times: terms
containing $a$ were then of a larger order of magnitude and produced a 
large growth only restricted by our boundary conditions at $t=t_e$.
Indeed, this leads us to an important conclusion: if $a=1$, we have equipartition
of magnetic field and radiative energy densities. This corresponds to an
equivalent-to-present field strength of 3 $\mu G$. Thus $a=10^{-5}$
would correspond to a present field strength of $10^{-8}G$.
Fields as low as equivalent-to-present $10^{-8}G$ are able to
affect inhomogeneities in the time interval considered. If they are
now measured to be higher than this, some amplification or dynamo mechanism
must have taken place after recombination. Magnetic fields which are
able to affect the small and intermediate scale inhomogeneities are 
also of the order of $10^{-8} G$ (equivalent
present values).

In Figures 2 and 3, we plot our results for small scale flux tubes, with
$\sigma =0.3$. Figure 2 shows the time evolution of the maximum perturbation
at the flux tube axis for $a = 10^{-5}$, taking two different initial boundary
conditions: isocurvature and inhomogeneity. We see that both initial conditions
give the same results except for very early times. Figure 3 shows the
time evolution of the inhomogeneity profile. It remains essentially
gaussian throughout the whole time period, increasing more rapidly in the recent
half time period.

\begin{figure}
\caption[]{Time evolution of the value of $\delta$ at the centre of the
filament, for $\sigma =0.3$ and $a= 10^{-5}$. Curve ¨0¨ for $\delta(t_0)
=0$. Curve -X for $\delta(t_0)=-X$}
\end{figure}
\begin{figure}
\caption[]{Time evolution of the filamentary inhomogeneity profile for
$\sigma =0.3$ and $a=10^{-5}$ for the boundary condition $\delta(t_0)=-X$.
The parameter characterizing the different curves is a time parameter}
\end{figure}

In Figures 4, 5, 6, we plot the results obtained for intermediate scale
flux tubes, with $\sigma=1$. Figure 4 shows the time evolution of the
maximum perturbation for $a= 10^{-4}$, $a=10^{-5}$ and for both isocurvature
$(\delta (t_0) =-X)$ and homogeneity $(\delta (t_0)=0)$ initial  conditions.
The first one provides curves without a short initial decrease, which does
not seem to be very realistic. For small magnetic fields we see again
that the growth is faster in the last part of the time considered. For large
magnetic fields the situation is more or less reversed. Figure 5 shows
the time evolution of the inhomogeneity profile, for moderate magnetic field
strengths, $a=10^{-5}$, and Figure 6 for higher strengths. The latter 
shows significant departures from the gaussian profiles.
\begin{figure}
\caption[]{Time evolution of the value of $\delta$ at the centre of the 
filament, for $\sigma=1$. Curve a: $a =10^{-5}$, $\delta(t_0)=0$;
curve b: $a=10^{-5}$, $\delta(t_0)=-X$; curve c: $a=10^{-4}$, $\delta(t_0)
=0$; curve d: $a=10^{-4}$, $\delta(t_0)=-X$}
\end{figure}
\begin{figure}
\caption[]{Time evolution of the filamentary inhomogeneity profile for
$\sigma =1$ and $a=10^{-5}$ for the boundary condition $\delta (t_0)=
-X$. The parameter characterizing the different curves is a time
parameter}
\end{figure}
\begin{figure}
\caption[]{Time evolution of the filamentary inhomogeneity profile for 
$\sigma=1$ and $a=10^{-4}$ for the boundary condition $\delta(t_0)=-X$.
The parameter characterizing the different curves is a time parameter}
\end{figure}

\section{Large scale flux tubes}

In our basic equation (2) the last two terms have 
orders of magnitude of $\delta /\sigma^2$ and $a/\sigma^2$, respectively. They 
can therefore be ignored when $\sigma$ is very large. This fact is
important for two reasons. From the integration point of view, if the laplacian
term is negligible, the equation is no longer elliptic. It can be treated
analytically and what is most noticeable is that it can be integrated
as an initial-value problem, thus restoring the prediction ability of
the equation. We do not need to assume the shape and magnitude of the 
inhomogeneity before the acoustic regime period. The other reason is that
$\sigma \gg 1$ probably represents a more interesting case as dissipative
effects may wipe out small scale inhomogeneities after the time epoch
considered in this paper.

The analytical solution now becomes
\begin{center}
  $\delta =-X +c_1 t +{c_2 \over t}$
\end{center}
where $c_1$ and $c_2$ are integration constants, which may be determined
with the boundary conditions. We have several possibilities which
should be discussed.

Suppose first that we are considering the time interval [0,1]. For 
$t=0$, we have $\delta =\infty$ unless $c_2 =0$. But then $\delta
(t=0)=-X$. Then only isocurvature would be a valid initial condition,
and $\delta =-X +c_1 t$. However, we must avoid $t=0$ (Big-Bang) and
begin with $t=t_0$, very small but non-vanishing. Also, in our numerical
outputs we have carried out the integration since $t=t_0 > 0$. We
have four options to determine $c_1$ and $c_2$:

1) Boundary value and homogeneity. We assume $\delta (t_0)=0$ and
$\delta (t=1)=\epsilon X$. The solution is then
\begin{center}
  $\delta =-X +X(1+\epsilon )t +{{Xt_0}\over t}$
\end{center}
As $t_0$ is low, this basically represents a linear growth.

2) Boundary value and isocurvature. We assume $\delta (t_0)=-X$, 
$\delta (1) =\epsilon X$
\begin{center}
  $\delta =-X +X(1+\epsilon)t +{{X(1+\epsilon)t_0^2} \over t}$
\end{center}
where the last term is negligible. We again obtain a quasi-linear growth.
As it was also numerically obtained, the conditions of homogeneity and isocurvature
provide the same results, except in the beginning.

3) Initial value and homogeneity. We assume $\dot{\delta}(t_0)=0$ and
obtain
\begin{center}
  $\delta =-X +{1 \over 2} {X \over t_0}t + {1 \over 2}X {t_0 \over t}$
\end{center}
which also represents a nearly linear growth, but now the rate of
growth $X/2t_0$ is larger than the $X(1+\epsilon)$ in the two previous cases.
In the end, we obtain $\delta (1)=X \left( -1+{1\over t_0}\right)
\approx X/t_0$, i.e. the inhomogeneity grows $1/t_0$-fold, where
$t_0$ could be considered the magnetogenesis time, even if we adopted
in the numerical integration in the previous section $t_0 = 0.01$.
This third possibility has the best prediction ability.

4) Initial value and isocurvature. We assume $\delta (t_0)=-X$ and
$\dot{\delta}(t_0)=0$. In this case, we obtain $c_1=c_2=0$ and
therefore
\begin{center}
  $\delta =-X$
\end{center}
is a constant. No evolution is to be expected.

With the exception of the fourth, these possibilities all, basically, lead
to the result already obtained for small scale flux tubes:
the growth is more or less linear. The growth is plotted in Figure 7
for the four possibilities mentioned above.
\begin{figure}
\caption[]{ Time evolution of the value of $\delta$ at the centre of the 
filament, for $\sigma \gg 1$ calculated with $\epsilon =10$, $t_0 =0.01$
and $X_{max}=10^{-5}$. Curve a: Boundary value and homogeneity; Curve
b: Boundary value and isocurvature; Curve c: Initial value and homogeneity;
Curve d: Initial value and isocurvature. Curves b and c  coincide 
except for small values of $t$, where curve b gives slightly higher 
values}
\end{figure}

There is another analytical integration in another approach. Suppose
that $\sigma \gg 1$ and that $a \gg \delta$, so that only the laplacian
term is negligible. The analytical solution becomes
\begin{center}
  $\delta= (\ln{t})t \hskip 0.5mm {{2a}\over \sigma^2} e^{-{r^2 \over \sigma^2}}
          \left({r^2 \over \sigma^2} -2 \right)$
\end{center}
but this solution does not represent a realistic case: $\delta$ has
a maximum at $t=e^{-1}$, vanishes at $t=1$, and what is more important,
$\delta$ increases until it reaches values comparable to $a$, in contradiction
with the initial assumption.

\section{Conclusions}
Primordial magnetic flux tubes are able to produce filamentary inhomogeneities
of density which grow more or less linearly throughout the epoch considered
in this paper. This epoch has been restricted from annihilation to just
before the acoustic epoch, from $z \approx 10^8$ to $\approx 10^5$, although the
calculations may account for the evolution in epochs earlier than this
time interval.

We have restricted ourselves to a particular case: that of a flux tube,
or more specifically to the field configuration given by equation (1). This
choice was in part due to the frequent observation of flux tubes in other
cosmic ionized systems and to the fact that they are suggested by the 
material filaments often observed in the large scale structure. It is
also a very simple symmetric structure defined with only one coordinate. 
However this particular choice restricts the generality of our results and
other field configurations are possible.

However, flux tubes, or at least structures as defined by equation (1),
constitute a rather general case if we consider a universe with no mean
magnetic field and in which the fluctuating field is made up of characteristic
cells with a coherent internal field orientation, but having no "a priori"
relation with the orientation of the field in adjacent cells. Suppose
a coherence cell in which the field can be represented by (0,0,$B_z$),
with $z$ clearly being the direction of the field within the cell. We look
for a function $B_z(x)$, i.e. when we leave the cell following a perpendicular
direction to the field in the cell. Before encountering another coherence cell,
$B_z(x)$ would vanish. In the opposite direction (-x) we would have the
same function. At the centre, we would have a maximum of $B_z$, with
$(\partial B_z/\partial x)_{x=0} =0$, avoiding a discontinuity
in the second derivative. If we adopt axisymmetry, we conclude that
$B_z(r)$ can reasonably be assumed to be a Gaussian, as specified by
our equation (1). It is true that flux tubes are long structures and the 
above argument does not consider the length of the structure defined
by (1). Along $z$, the field can be independent of $z$ and therefore represented
by (0,0,$B_z(r)$) within a z-length, which is considered long in this paper,
but with an unimportant and unspecified value. Therefore, even if we consider
a particular magnetic field configuration as our basic structure, our results
remain rather general.

After the epoch considered in this paper, other effects, such as damping of
small diameter filamentary structures prior to recombination, non-linear
growth after recombination and mechanisms amplifying magnetic fields in
recent pregalactic and galactic epochs, will complicate this simple
picture, but this is beyond our objectives.

Our work is restricted to an unobservable epoch, which means that our results
cannot strictly be compared with observations. Its objective is to provide initial
conditions for other models devoted to more recent epochs, the whole history
of cosmological magnetism being a huge task, which cannot be undertaken by a 
single model. Nevertheless, it is unlikely that large filamentary structures have disappeared
in more recent epochs, so that our results may provide an explanation for
presently observed large scale structures. The implications of this model
in their interpretation are in part, considered in Paper III.
It is really to be expected that large structures remain unaffected by
complex processes after equality. This is justified as follows:

Damping of cosmic magnetic fields has been considered by Jedamzik,
Katalinic and Olinto (1997) introducing viscosity, bulk viscosity and
heat conduction. After neutrino decoupling the main damping mechanism is
photon diffusion, which only affects structures up to the Silk mass, 
about $10^{13} M_\odot$.

When considering the growth of unmagnetized structures, it is assumed that
non-linear effects are important only at smaller scales, up to $\sim$10 Mpc,
and this limit probably remains valid when magnetic fields are taken into 
account. In practice, this limit corresponds to the scale at which the rms
galaxy fluctuations are unity, and is therefore independent of the involved
forces. This issue has been considered by Kim, Olinto and Rosner (1996).

Finally, we must take into account specific mechanisms of magnetic field
creation and amplification in recent times. A large variety 
have been proposed (Zwibel, 1988; Pudritz and Silk, 1989; Harrison, 1973;
Tajima et al., 1992; Lech and Chiba, 1995, and others); See also the reviews by Rees (1987)
and Kronberg (1994). The important fact is that these mechanisms induce
small scale magnetic fields, smaller than a few Mpc. As far as we are aware,
no mechanism for producing magnetic fields at scales larger than a few
Mpc have been proposed for post-Recombination mechanisms.

After Equality, some mechanisms erase pre-existing magnetic fields, and
others amplify them in a complicated way, thus modifying the pre-Equality
magnetic fields considered in this paper, although these mechanisms only
affect the small scale structures. The evolution of large scale structures could be
described by the formulae in Appendix B into paper I,
i.e., the structures are maintained, simply growing with the expansion.
Our model therefore, constitutes a tool to interpret present large scale
structures.

It is a well established observational fact that the large-scale structure
of the Universe is very rich in filaments (Gregory and Thomson, 1978;
Oort, 1983; de Lapparent et al. 1986 and others. See for instance the review by
Einasto, 1992) being more abundant than two-dimensional
sheets (Shaty, Sahni and Einasto, 1992). They can play an important
role in the formation of clusters (West, Jones and Forman, 1995).

The existence of large-scale filaments is currently accounted for by
other hypotheses, but it is here suggested that primordial magnetic
flux tubes constitute an additional alternative, or at least, a mechanism
reinforcing other gravitational effects. Filaments are associated
with magnetic fields in many astrophysical systems, such as the Sun and the
interstellar medium, and we now see that this association can be extended
to large-scale filamentary structures in the Universe.

The best studied large-scale filamentary structure is the Coma-A1367
supercluster, which is itself elongate and extended towards the 
Hercules supercluster. Its diameter is about 10 Mpc, thus constituting
a large scale inhomogeneity in the sense considered here (i.e. $\gg$
0.28 Mpc; $\gg$ 1 in the units defined above). Its length can be very
large (Batuski and Burns, 1985). The distribution of early type
galaxies is particularly thick (Doi et al., 1995). 

As observed random velocities of groups and clusters with respect to the
filament structure are relatively small ($\sim 100 Km \hskip 0.5mm s^{-1}$; Tully, 1982)
the observed distribution of galaxies reflects its distribution when the whole
structure was formed. The evolution of the filament and of the network
of filaments it belongs to, has evolved very little.

Rather interestingly, the magnetic field strength has been measured in this
supercluster. In a region well outside the coma cluster in the direction toward
A1367, Kim et al. (1989) observed a bridge of synchrotron emission with
the same direction, of about 0.3-0.6 $\mu G$, a large value for an extracluster
region. In the Coma cluster core region it is even larger, of the order
of 1.7 $\mu G$ (Kim et al. 1990). Radio observations of the Coma cluster
and its vicinity at different frequencies have been reported by Kim et al
(1994) and Kim (1994). It would be very interesting to determine whether the
direction of the magnetic field coincides with the NE-SW direction, which
is that of the huge filament. This coincidence would be in noticeable 
agreement with the model here suggested.

\end{document}